# DV-ARPA: Data Variety Aware Resource Provisioning for Big Data Processing in Accumulative Applications

*Hossein Ahmadvand and Fouzhan Foroutan*
*ahmadvand@ce.sharif.edu*
*Department of Computer Engineering Sharif University of Technology*

*Abstract* – **In Cloud Computing, the resource provisioning approach used has a great impact on the processing cost, especially when it is used for Big Data processing. Due to data variety, the performance of virtual machines (VM) may differ based on the contents of the data blocks. Data variety-oblivious allocation causes a reduction in the performance of VMs and increases the processing cost. Thus, it is possible to reduce the total cost of the job by matching the VMs with the given data blocks. We use a data-variety-aware resource allocation approach to reduce the processing cost of the considered job. For this issue, we divide the input data into some data blocks. We define the "significance" of each data block and based on it we choose the appropriate VMs to reduce the cost. For detecting the significance of each data portion, we use a specific sampling method. This approach is applicable to accumulative applications. We use some well-known benchmarks and configured servers for our evaluations. Based on the results, our provisioning approach improves the processing cost, up to 35% compared to other approaches.**

1. INTRODUCTION

Choosing a suitable server configuration in big data processing is an important issue for cloud providers and users. In this paper, we focus on the impact of data variety on the performance of VMs to decrease the cost of cloud computing for Big Data processing. Our previous observations have shown that various portions have a different impact on results due to data variety [1]. Based on the results reported in [1], changes in the sequence of input data can improve the speed of generating the result. We also have shown that changing the processing infrastructure due to data variety can improve the quality of Result [2]. A variety aware approximation approach based on sampling is offered in [3]. In continuation of these researches, we consider Service Level Objectives (SLO) such as Preferred Finishing Time in the current paper. Cloud providers and users can use this approach to increase their benefit of using cloud computing for Big Data processing.

As the first step of our solution, we need an approach to define the effect of each data portion on the final outcome. So, we define a certain *Significance* measure for each application. Significance measure must be defined in a way that shows the progression of computation. For example, in WordCount application, the number of words is the significance measure. In the second part of our solution, we divide input data into some same size portions. As the next step, we determine the significance of each portion using sampling with a 95% confidence interval and a 5% margin of error. We categorize data portions into some Data Types based on their significances. Third, we design an approach to detect the cost/time efficient server for each Data Type. Finally, we allocate the portions to the servers in a way that reduces the processing cost. Fig. 1 shows our approach in an abstract view. Based on the results, our approach improves the cost of processing while the SLOs (Service Level Objectives) associated with the workload are met. Our evaluations have shown that our approach improves the cost of processing up to 35%.

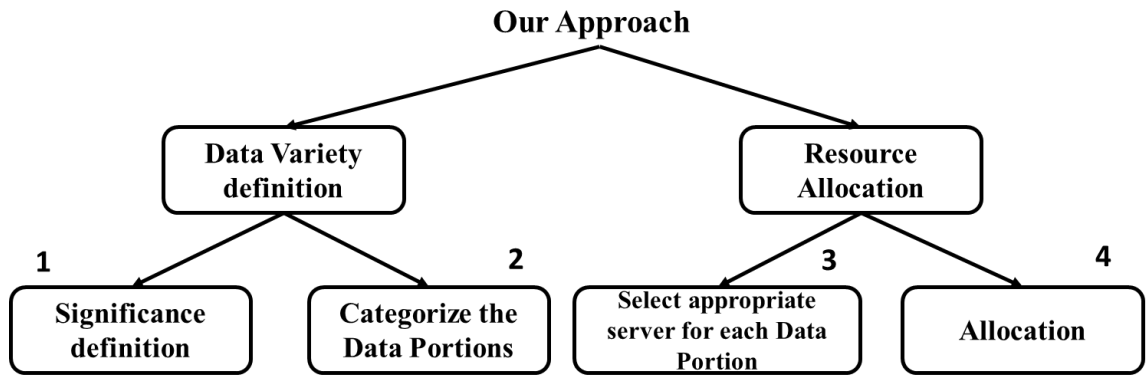

Fig. 1. Our approach

In the evaluation section, we have used apache Spark version 2.0 on Ubuntu12.04 as the framework of our experiments. The specification and price of servers have obtained from Amazon EC2 [4]. This means that we configure our servers based on the Amazon EC2 model.

**Contribution.** Our main contributions in this paper are listed below:
- First, we present an approach for decreasing processing costs using Data-Variety and management of infrastructure. We improve processing costs by assigning data portions to appropriate servers.
- Second, we evaluate our data-variety-aware approach with several applications and various datasets. The experimental results show that our data-variety-aware approach can surpass data-variety-oblivious approaches and improves processing costs. We use seven public datasets and six well-known applications in our evaluations.
- Third, our approach is simple and applicable for users and cloud providers.

The rest of the paper is organized as follows. The related works are presented in Section 1.1. Section 1.2 presents a motivational example. Problem definition and the proposed approach are explained in Section 3. We evaluate our work with a range of applications and datasets in Section 4. Finally, we present conclusions and future works.

1.1 RELATED WORKS

In this subsection, we have presented previous related works in our research area. The previous works are divided into some categories:

**Variety.** The authors in [5] have presented techniques and tools to detect the most important portions of code and variables in output quality. They reduce the processing energy by using this approach. We also have used data variety for increasing the Quality of results in case of lack of time and processing cost [2]. The authors in [6] have declared the definition of 3Vs of Big Data and have used them for meeting SLA in the software system in Big Data processing. We also focus on data variety to meet the SLA in the current paper.

The authors in [7] have analyzed the behavior of Big Data processing at the microarchitecture level. This paper has focused on the architecture of Big Data processing to reduce the energy consumption of Big Data processing. Their platform is available for cloud providers and users. In this area, paper [8] has presented a distributed framework for big Data processing which is specifically used in smart city data. The authors in [9] have presented a framework to help the programmers for Big Data processing. We also have presented a framework to help the users and programmers to process their Big Data at a lower price.

**Framework.** Similar to our work, the authors in [10] and [11] have presented an approach for assigning queries to the suitable processing infrastructure. They have defined the Quality of Result based on their objective function. In [12], the authors have focused on finding a suitable configuration for processing queries to increase the throughput. They have used some meta-heuristic approaches for their objective. They have considered the "number of tuples that can be processed in a unit of time" as the measure of throughput. We also have defined a measure to determine the progression of each application process. They have not considered data variety in their definition. Paper [13] presented an approach for assigning queries to existing VMs. The authors have reduced the load on expensive servers and improved resource costs by executing queries on VMs with lower costs, first. They have predicted that the market share of cloud providers will be increased using this approach. The authors in [14] have shown the impact of different Hadoop clusters on the performance of the counter-based detector. They have used this approach for detecting DDoS attacks with URL counting. The authors in [15] have used reliable hardware for the most significant tasks and the unreliable hardware for the least significant ones. We also have presented an approach to increase the Quality of Result by using data variety in case of budget and time constraints [2]. The authors in [16] focused on the reduction of the energy cost in clouds. They have offered a pricing policy to cloud providers for increasing the profit. The authors in [17] present an approach to select the most cost-effective configuration in a way that resource costs are minimized while the Service Level Agreements (SLAs) associated with the workload are met. We also have offered an approach to improve the Quality of Result while meeting the time and budget limitations of the problem.

**Scheduling.** The main contribution of [18] is scheduling with respect to deadline and budget. The authors in this article have optimized the scheduling of MapReduce jobs at the task level. In this paper, the authors have focused on two aspects of scheduling optimization. They have considered fixed time or budget and optimized the other parameter. The authors have presented the pipelined version of Hadoop in paper [19] and they named it Hadoop Online Prototype (HOP). They have increased the performance of this framework by choosing a desirable split of data for feeding the pipeline. We have focused on selecting suitable data portions to be processed by servers. Authors in [20] consider the characteristics of data for resource allocation. In this article, input data consists of text, image, video, and audio. The authors present an approach for sampling and estimating the volume and velocity of the various type of input data. Resources are allocated to input data based on its functionality and requirement. In our paper, we consider aggregate applications and evaluate the impact of the variety of input data on the result. We also consider the Variety that is one of the 4Vs of Big Data. Authors in [20] showed that the allocation should be changed whenever data characteristics change in terms of type, volume, and velocity. Similar to our work, the definition of progress interval by data sampling is the main concept in [21]. We also have used progressive computing in previous works [1].

**Accuracy of the processing.** In paper [22] approximation is considered as a solution to increase the performance of Big Data processing. Authors have presented their approach by changing the MapReduce structure and architecture of Hadoop. They have achieved acceptable results by processing some parts of data. In our research, we have focused on the significance variety in various parts of data. We also have presented a novel approach to increase the accuracy of approximate computing in the presence of data variety in [3]. In this paper, we offer an approach to determine the size of data portion and samples. The mentioned approach has achieved an acceptable result in compare to other existing approaches. We also have presented a data variety approximation approach in [3].

No one has yet used Data-Variety for scheduling of Big Data jobs with respect to budget and preferred finishing time constraints to increase the benefit of cloud computing for Big Data processing. As another contribution, we present a detection approach to assign an efficient configuration for each part of data that is ignored in previous researches.

1.2 MOTIVATION

In this subsection, we have presented some pieces of evidence that motivated us to continue the research. We divide the input data into some same size portions and show that these portions have a different impact on the final result. Fig. 2 shows this diversity of various portions' impact in terms of significance and processing cost for WordCount application. In this example, we consider 4GB data of the IMDB dataset and divide it into 4 portions.

Fig. 2.A shows that if we use heterogeneous servers instead of homogenous servers, we can manage the processing time. As Fig. 2.B shows, the first portion is the most significant one. This portion generates the main part of the result. We processed the data portions with a one-type server and showed the processing cost of each portion in Fig. 2.C. In Fig 2.D, we use a higher configuration server for the first portion. As Fig 2.D shows, the processing time of the first portion is increased and we achieve the main part of the result faster. By processing the most significant data portions with higher configured servers, we can use our budget in a more efficient way in case of time and budget constraints.

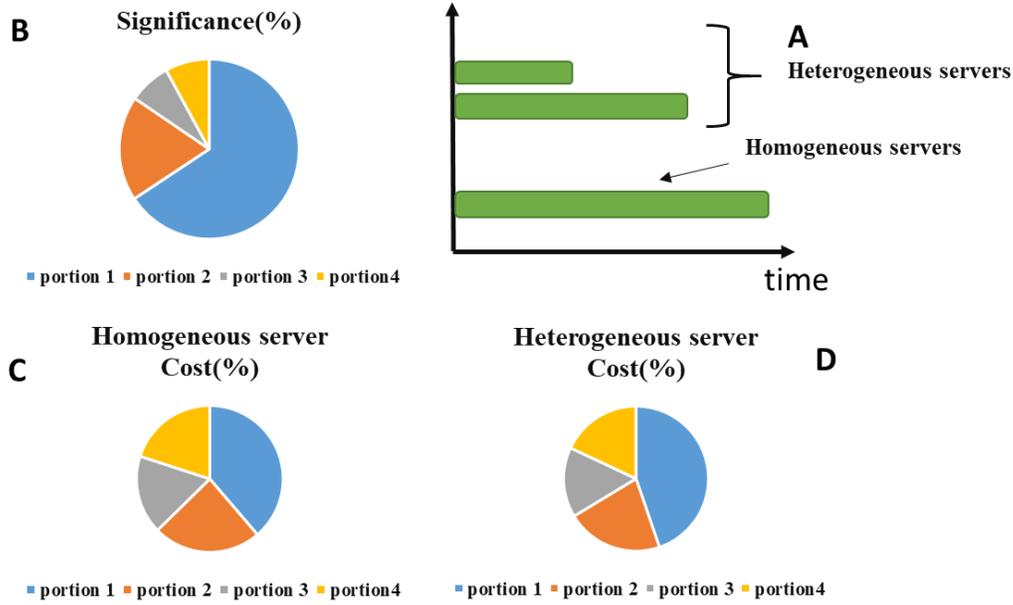

Fig. 2. motivation for our work

This motivational example shows that data variety has an important effect on the processing time and cost. So, we have continued the research to achieve an approach to overcome the impact of data variety.

In this paper, we analyze *accumulative applications*. In this kind of applications,
- ✓ Various parts of data are independent of each other
- ✓ The final result is generated by concatenating the partial results
- ✓ The application is based on the accumulative operation such as sum, average, Min, Max

In this kind of application, Data Portions that are more significant consume more processing costs. For example, in WordCount application, data portions with more words, require more processing units to run.

On the other side, in graph applications, input data have a dependency and the final result cannot be generated by concatenating the partial results. So, our approach is not applicable to graph applications.

2. METHODS

In this section, we explain the Problem and discuss our methodology to solve it. First of all, subsection *A* describes the mathematical definition of our problem. We present the notation of our algorithm in Table 1. After that, we present the problem statement and the problem formulation. Subsection *B* presents our algorithm and then ***Complexity Analysis*** is presented.

### A. Problem Definition

In this paper, we reduce the Processing Cost and meet the Preferred Finishing Time. We use some notations for showing the problem definition. **TABLE 1** presents these notations. We have used this notation for mathematical description.

### Problem Statement.

We want to minimize the Processing Cost and meet the Preferred Finishing Time. For this goal, we should assign each data portion to the appropriate server. We should consider the processing time/cost of each data portion of the available servers.

TABLE 1. Notation of our algorithm

| Notation | Description |
|---|---|
| MSDT | Most Significant Data Type |
| LSDT | Least Significant Data Type |
| MeSDT | Medium Significant Data Type |
| CMSDT | Processing Cost of MSDT |
| CLSDT | Processing Cost of LSDT |
| CMeSDT | Processing Cost of MeSDT |
| PFT | Preferred Finishing Time |
| FT | Finishing Time |
| DP | Data Portion |
| DT | Data Type |
| EF | Efficiency parameter |
| CPP | Cost Per Performance |
| PT | Processing Time |
| CPTU(1…n) | Cost Per Time Unit |
| PC | Processing Cost |
| ST | Server Type |
| TCP | Time Critical Path |
| ES (1...p) | List of Efficient Servers |
| NS | Number of Servers |
| NP | Number of Portions |

***Problem Formulation.***
The objective function, which should be minimized, is the total Processing Cost(PC), and the Preferred Finishing Time (PFT) is the constraint of our work.

$$Min(PC) \qquad (1)$$

Subject to:

$$FT < PFT \qquad (2)$$

Formula (1) presents our objective in this work. Formula (2) presents the constraint of our work. Finishing Time (FT) must have a lower value in comparison with PFT.
Formula (3) shows the total processing cost. Total Processing Cost consists of the processing cost of each Data Type.

$$Cost = CPTU * \sum PT \qquad (3)$$

Formula (4) presents the performance of processing. The numerator is the result and the denominator is the sum of Processing Time. In other words, the performance of processing is the speed of progressive processing.

$$Performance = (\sum Result) / \sum PT \qquad (4)$$

Formula 5 defines the result of processing. The result is the sum of all significances. Significance Measure determines the number of results generated. On the other hand, a certain data block is more important if it has more significance. For example, in WordCount application, the number of words in each data portion is the significance value.

$$Result = \sum Significance \qquad (5)$$

We also define the EF parameter to define the effect of Data Types on the final outcome. Formula (6) shows the EF definition. As Fig.3 shows, we divide the input data into 3 Data Types based on EF. In other words, EF shows the ratio of significance to the volume of the considered data portion.

$$EF_i = (significance_i / \sum significance) / (volume_i / \sum volume) \quad (6)$$

We define a parameter CPP to detect the cost-effective server for each Data Type. Formula (7) presents the Cost Per Performance (CPP). We present the Cost definition in Formula 3 and the performance definition in formula 4.

$$Cost / Performance (CPP) = CPTU * (\sum PT)^2 / \sum Significance \quad (7)$$

- ✓ EF is used for categorizing the input data into some groups.
- ✓ CPP also is used to select the cost-effective Server Type for each Data Type.

### B. Our Algorithm.
We present a heuristic algorithm to solve this problem. In this heuristic, we categorize the input data into 3 types based on the EF and then we assign the Data Portions to the appropriate servers based on the CPP, DT, and ST.

*Algorithm*. We present algorithm 1 for our proposed approach. In the following, we discuss different parts of the algorithm.

| Algorithm1 |
|---|
| 1: Input: NS, NP, CPTU (1…n), PFT, ES |
| 2: output: PC |
| 3: divide DPs into 3 types (based on EF) |
| 4: estimate (CPP per DT and ST) |
| 5: Sort server Types based on CPP for each Data type |
| 6: select S1, S2, and S3 with minimum CPP for MSDT, MeSDT, and LSDT |
| 7: assign LSDT to S1, MeSDT to S2, MSDT to S3 |
| 8: estimate FT |
| 9: while (! meet PFT) |
| 10:   if (FT >PFT) |
| 11:     detect (TCP) |
| 12:     if (TCP= S1) |
| 13:       change S1 with a high configured server based on the CPP list |
| 14:     else if (TCP= S2) |
| 15:       change S2 with a high configured server based on the CPP list |
| 16:     else replace S3 with a more expensive server |
| 17: end while |

We present the initial values in line 1 and 2 of algorithm 1. In line 3 the input data is divided into 3 types. These types are MSDT, MeSDT, and LSDT.
We compute CPP for each Data Type and Server Type then select the server with better CPP to meet the SLO requirements (line 5). We assign Data Types to the servers (lines 6 to 8). If the Finishing Time has a bigger value compared to Preferred Finishing Time then detect the TCP (Time Critical Path). We detect the server that causes Time Critical Path and changes it with a higher configured server according to CPP.

*Complexity analysis.* The time complexity of our algorithm is of O (n + m) where n is the number of portions. The time complexity of Data Type definition for each Data Portion (line 4) is O(n). Line 10 to 16 is executed at most m times where m is the number of server types; thus, it is of O(m) complexity. Consequently, the overall complexity of the algorithm in terms of time is of O (n + m). For additional details, as Fig. 1 shows, our solution consists of two steps. In the first step, we divide the input data into some same size portions and define the significance of each portion. We use sampling in the calculation of significance of each data portion. We use *Cochran Sampling Technique* with a 95% confidence interval and a 5% error margin [23]. We define three Data Types based on the value of EF.

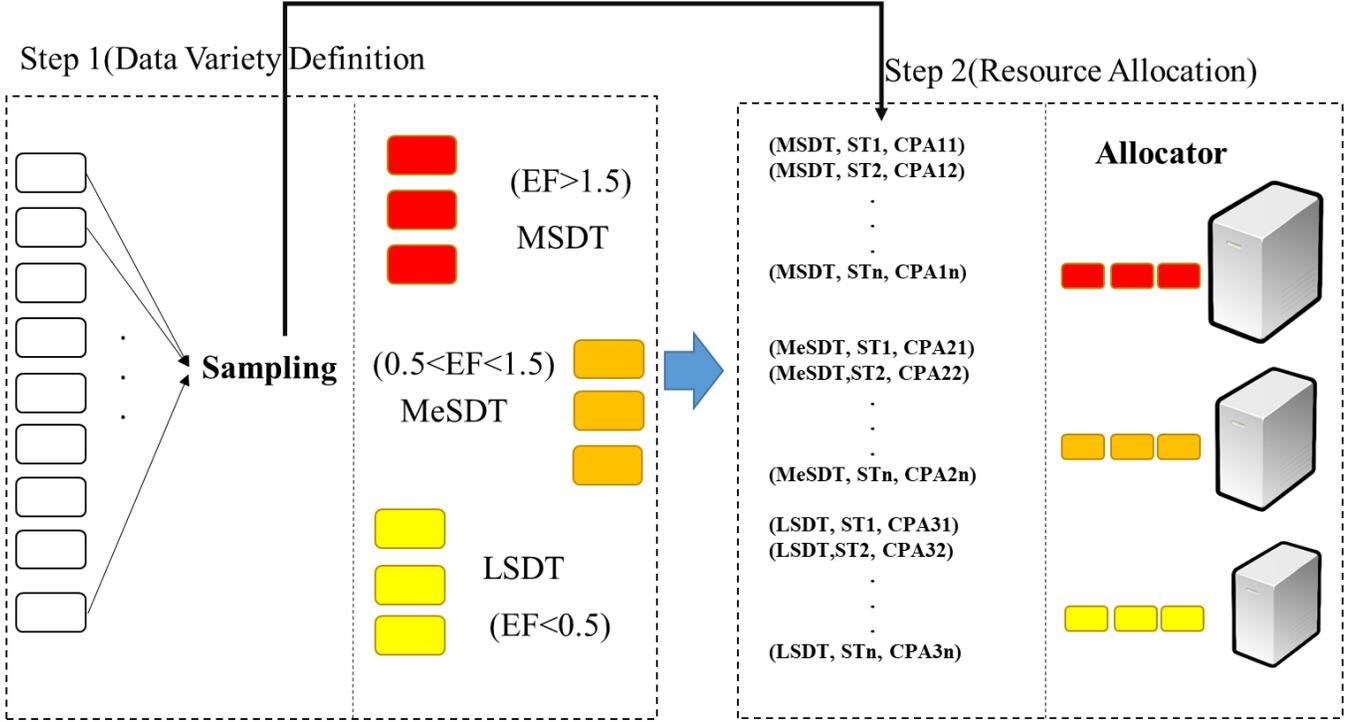

Fig. 3. Steps in our approach

$$PC = C_{MSDT} + C_{MeSDT} + C_{LSDT} \quad (8)$$

As formula (8), shows the total processing cost consists of the processing cost of each Data Types.

As Fig. 3 shows, in the first part of our approach, we divide the input data into some Data Types based on the EF. In the second step of our work, we assign Data Types to appropriate servers based on the CPP. For this goal, we compute CPP per Data type and Server type then sort the server types based on CPP for each Data Type. Then we assign the Data types to servers with minimum CPP.

3. RESULTS AND DISCUSSIONS

We evaluate our work by using the server configuration like Amazon EC2 configuration [4]. We have used Apache Spark version 2.0 on Ubuntu12.04 as the framework of experiments [24]. TABLE 2 shows the servers and configurations that we used. We have implemented these configurations on the infrastructure in our lab.

TABLE 2. Server configurations

| Name | Memory | CPU | Cost($/h) |
|---|---|---|---|
| S1 | 4 | 4 | 0.239 |
| S2 | 8 | 8 | 0.489 |
| S3 | 16 | 16 | 0.959 |
| S4 | 32 | 32 | 1.919 |
| S5 | 64 | 64 | 3.838 |

**Applications.** We use some well-known applications as our benchmarks [25]:
- *WordCount*. This application counts the number of words in the file.
- *Grep*. It searches and counts a pattern in a file.
- *InvertedIndex*. This application is an index data structure storing a mapping from content to its location in a database file.
- *Health*: This counts the number of volunteers with high blood pressure.
- *Investment*: This counts the value of the investment in certain States.
- *URL Counting*: This counts certain URLs in system logs.
- *AVG(TPC)*: The average of certain values in the dataset.

- *SUM(Amazon)*: SUM of reviewers' ranks.

We consider "*Significance Measure*" for each application that can be used for declaration of processing progress. We explain the Significance Measurement for each application like this:
- *WordCount*: The number of words existing in a certain data portion.
- *Grep*: The number of a specific phrase in a file.
- *InvertedIndex*: The size of the output index file.
- *Health:* The number of volunteers with high blood pressure.
- *Investment*: The value of the investment in certain State.
- *URL counting*: The number of a specific URL in a portion.
- *TPC and Amazon*: We also consider AVG for TPC-H datasets and SUM for Amazon datasets.

*Data Sources.* Real datasets are collected from four sources. We use Bootstrapping approach for generating enough volume of data [26]. TABLE 3 describes the datasets that we used for our evaluation:

TABLE 3. Datasets

| Name | Sources | Volume |
|---|---|---|
| IMDB | [27] | 500GB |
| Gutenberg | [28] | 500GB |
| Quotes | [29] | 500GB |
| Wikipedia | [25] | 500GB |
| MHEALTH Dataset | [30] | 2TB |
| Company Funding Records | [31] | 2TB |
| TPC-H(MAIL) | [32] | 2TB |
| TPC-H(SHIP) | [32] | 2TB |
| TPC-H(AIR) | [32] | 2TB |
| TPC-H(RAIL) | [32] | 2TB |
| TPC-H(TRUCK) | [32] | 2TB |
| Amazon (Music) | [33] | 2TB |
| Amazon (Books) | [33] | 2TB |
| Amazon (Movies) | [33] | 2TB |
| Amazon (Clothing) | [33] | 2TB |
| Amazon (Phones) | [33] | 2TB |

**SLOs Conditions.** We evaluate our proposed approach in two conditions. We explain these two conditions in TABLE 4. For each application, we present the PFT in each condition.

TABLE 4. Two forms of SLO

| applications | PFT(h) | |
| | Strict | Normal |
|---|---|---|
| WordCount | 10 | 11 |
| Grep | 5 | 6 |
| InvertedIndex | 2000 | 2200 |
| Health | 6 | 7 |
| Investment | 5 | 6 |
| URL counting | 6 | 7 |
| TPC-H(MAIL) | 5.5 | 6 |
| TPC-H(SHIP) | 5.5 | 6 |
| TPC-H(AIR) | 5.5 | 6 |
| TPC-H(RAIL) | 5.5 | 6 |
| TPC-H(TRUCK) | 5.5 | 6 |
| Amazon (Music) | 5.5 | 6 |
| Amazon (Books) | 5.5 | 6 |
| Amazon (Movies) | 5.5 | 6 |
| Amazon (Clothing) | 5.5 | 6 |
| Amazon (Phones) | 5.5 | 6 |

**Competitor Approaches.** We use three data-variety-oblivious approaches to compare with our work.
*WEAK.* Running all input data on the weak server (s1) with naïve MapReduce algorithm.
*MODERATE.* Running all input data on a moderate server (s2) with naïve MapReduce algorithm.
*STRONG.* Running all input data on strong and the most expensive server (s3) with naïve MapReduce algorithm.

3.1 RESULTS

Fig. 4 and Fig. 5 shows the processing time and cost in Normal condition. In Normal condition, our approach, Moderate and Strong can meet the SLOs. Our approach improves Processing Cost for WordCount, Grep, InvertedIndex, Health, URL-Counting and Investment by 30 %, 31%, 35%, 31%, 32% and 29% in compare to the Strong approach and 2%, 4%, 8%, 3%, 9% and 2% compared with the moderate approach. Fig. 6 and Fig. 7 shows processing time and processing cost in Strict condition. In Strict condition, only our approach and Strong can meet the SLOs. However, our approach saves more budget in compare to the Strong approach. Our approach decreases the Processing Cost in WordCount, Grep, InvertedIndex, Health, URL-counting, and Investment 18%, 27%, 13%, 18%, 23% and 17% compared with the Strong approach.

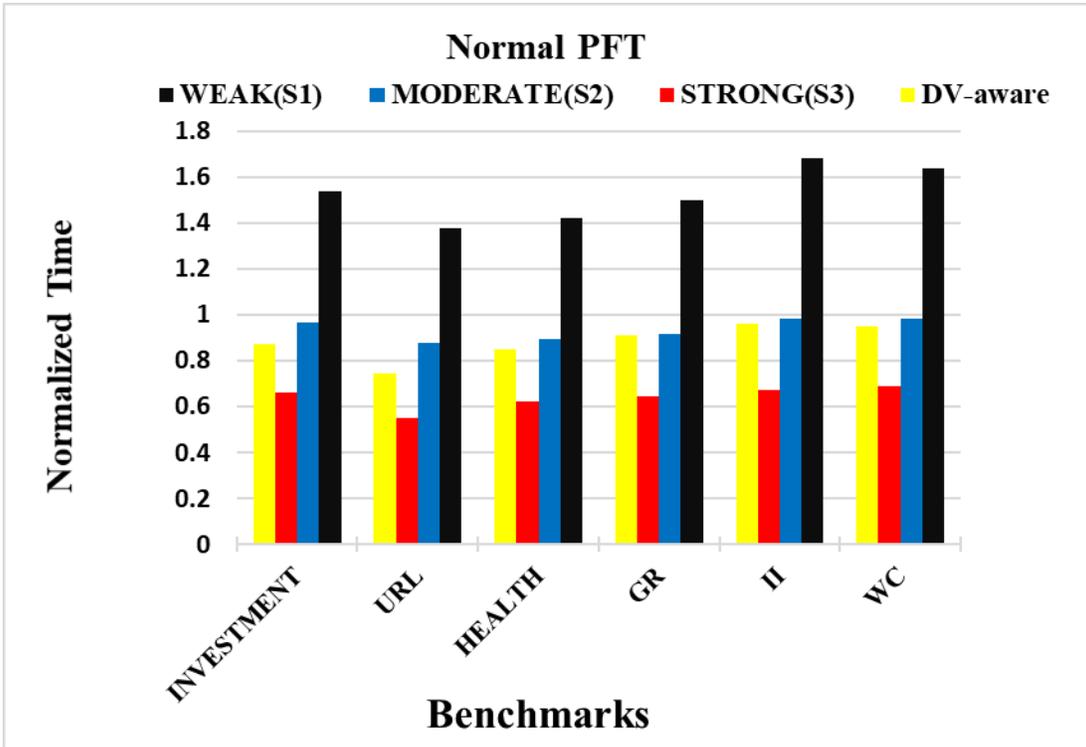

Fig. 4. Normalized Time in Normal SLO condition (Investment, URL, Health, Grep, Inverted Index and WordCount)

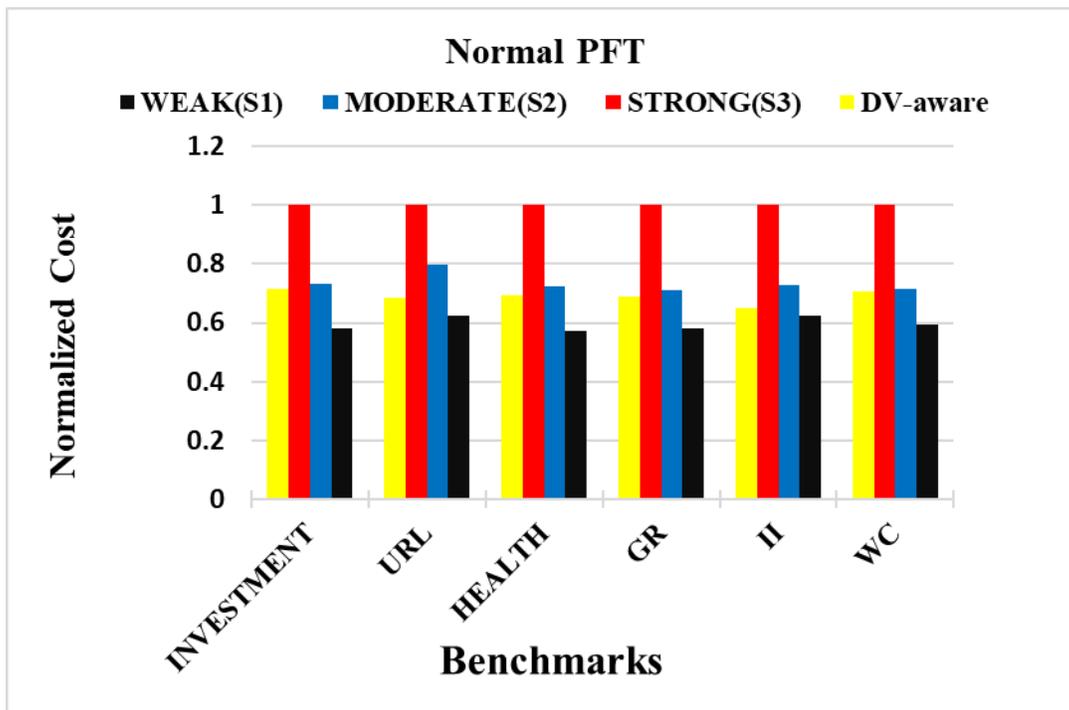

Fig. 5. Normalized Cost in Normal SLO condition (Investment, URL, Health, Grep, Inverted Index and WordCount)

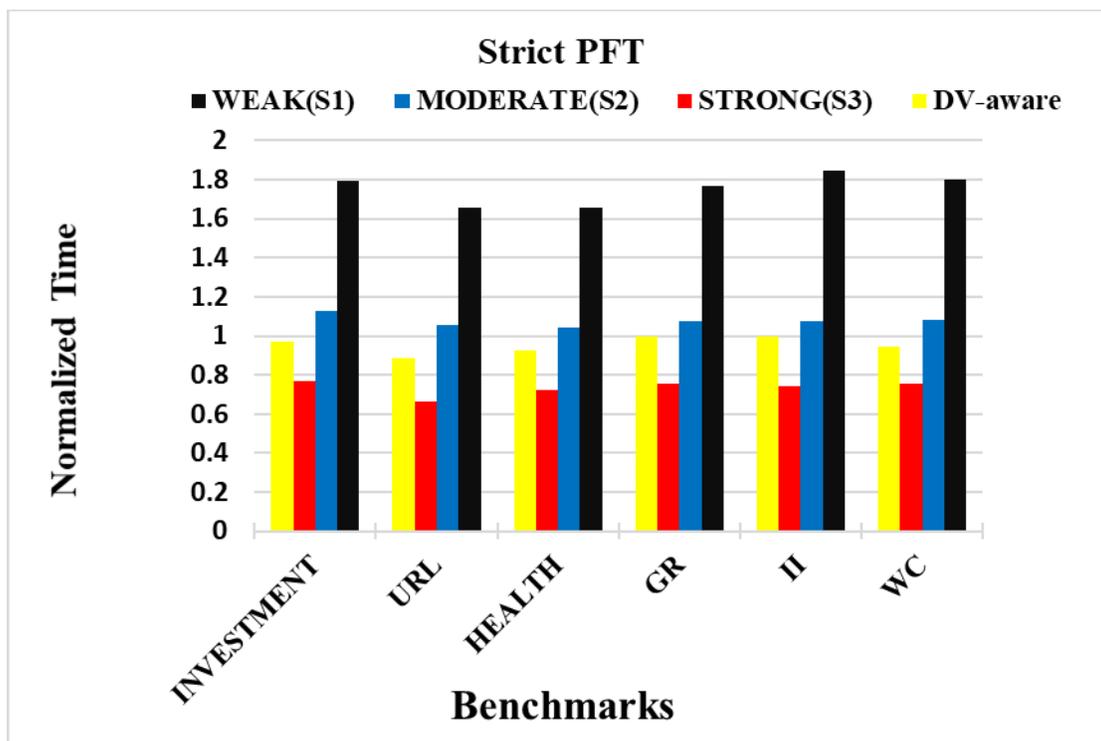

Fig. 6. Normalized Cost in Strict SLO condition (Investment, URL, Health, Grep, Inverted Index and WordCount)

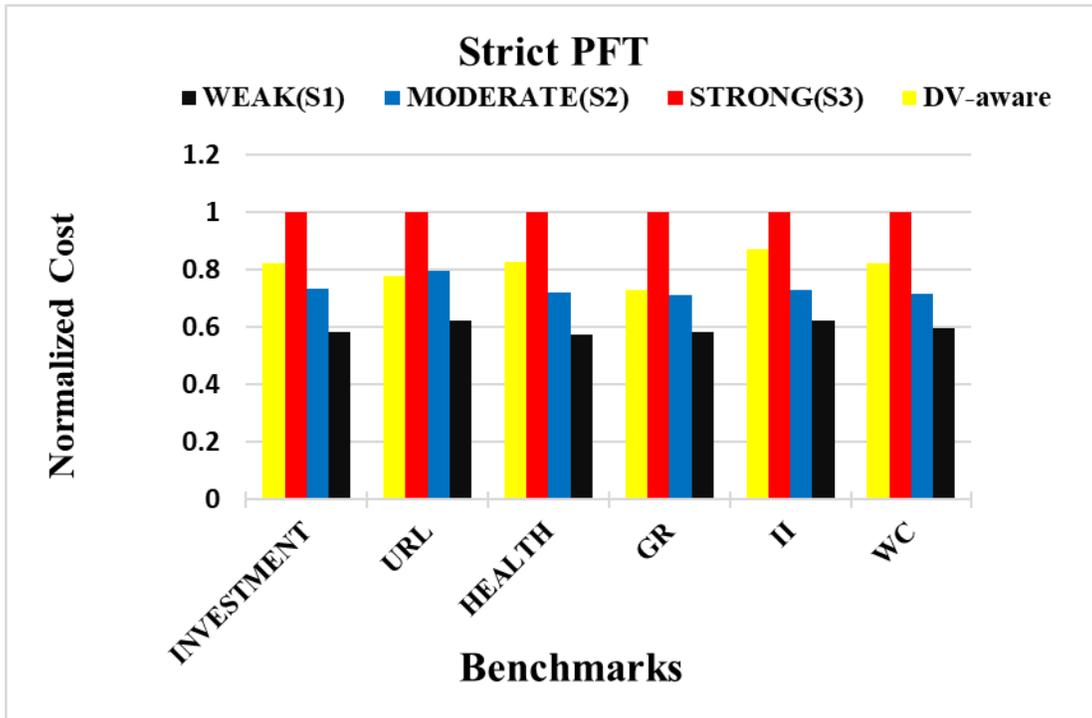

Fig. 7. Normalized Cost in Strict SLO condition (Investment, URL, Health, Grep, Inverted Index and WordCount)

**Fig. 8** and **Fig. 9** shows the processing time and cost in Normal condition for TPC-H benchmark. In Normal condition, our approach, Moderate and Strong can meet the SLOs. Our approach improves Processing Cost for TRUCK, RAIL, AIR, SHIP and MAIL by 35 %, 28%, 32%, 29% and 30% in compare to the Strong approach and 3%, 2%, 1%, 1% and 7% compared with the moderate approach. **Fig. 10** and **Fig. 11** shows processing time and processing cost in Strict condition. In Strict condition, only our approach and Strong can meet the SLOs. However, our approach saves more budget in compare to the Strong approach. Our approach decreases the Processing Cost in TRUCK, RAIL, AIR, SHIP and MAIL 26%, 17%, 22%, 26% and 24% compared with the Strong approach.

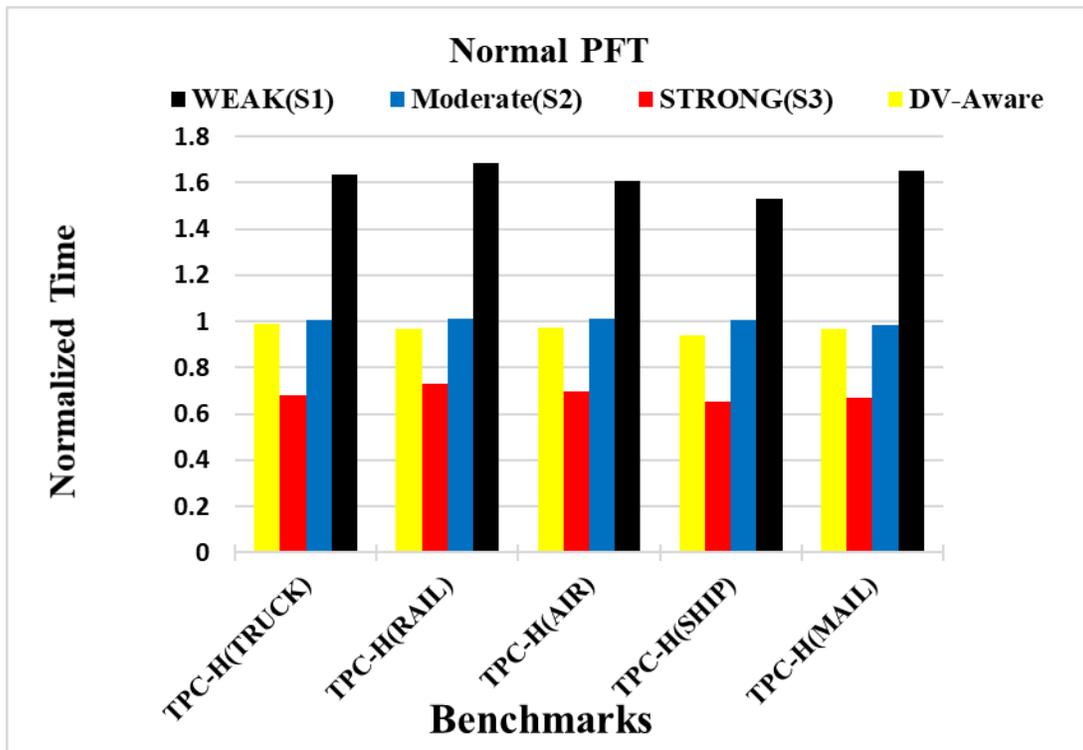

Fig. 8. Normalized Time in Normal SLO condition (TPC Benchmarks)

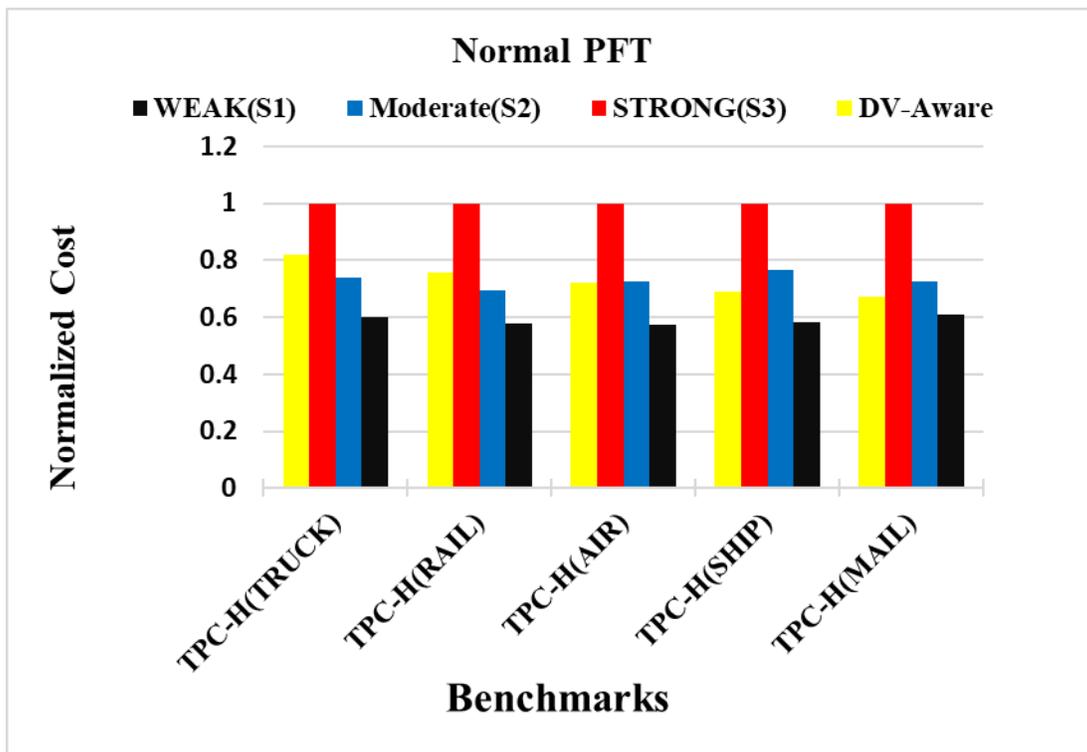

Fig. 9. Normalized Cost in Normal SLO condition (TPC Benchmarks)

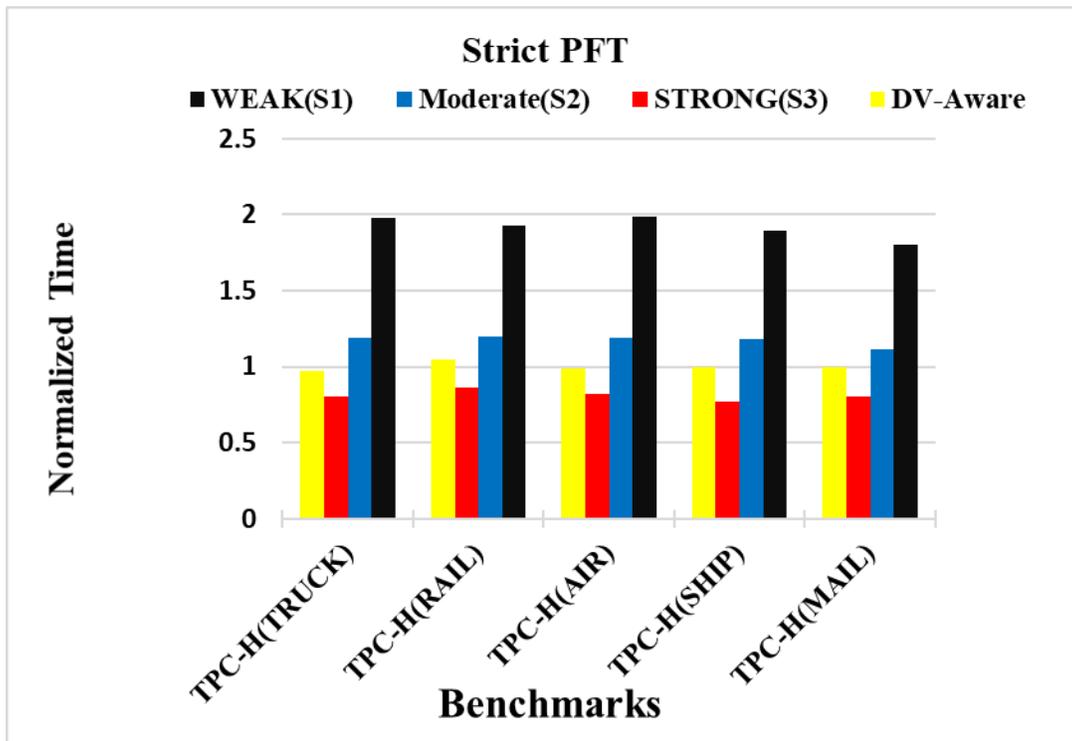

Fig. 10. Normalized Time in Strict SLO condition (TPC Benchmarks)

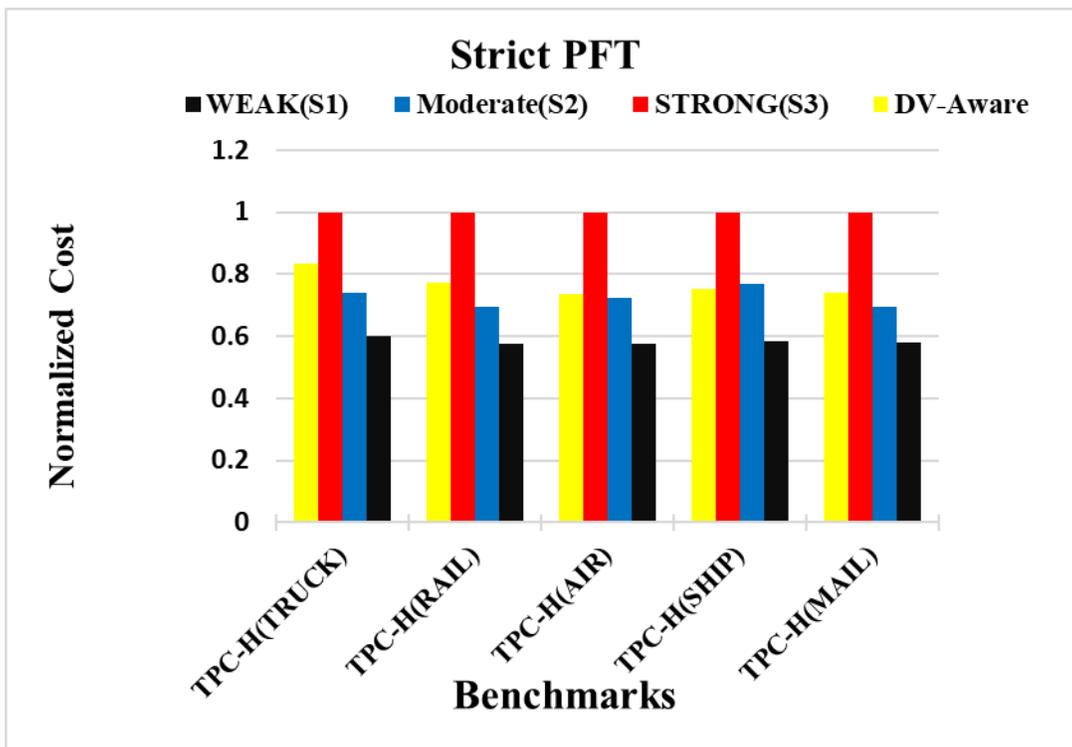

Fig. 11. Normalized Cost in Strict SLO condition (TPC Benchmarks)

**Fig. 12** and **Fig. 13** shows the processing time and cost in Normal condition for Amazon benchmark. In Normal condition, our approach, Moderate and Strong can meet the SLOs. Our approach improves Processing Cost for Music, Books, Movies, Clothing and Phones by 29 %, 25%, 32%, 29% and 18% in compare to the Strong approach and 4%, 4%, 2%, 2% and 2% compared with the moderate approach. **Fig. 14** and **Fig. 15** shows processing time and processing cost in Strict condition. In Strict condition, only our approach and Strong can meet the SLOs. However, our approach saves more budget in compare to the Strong approach. Our approach decreases the Processing Cost in Music, Books, Movies, Clothing and Phones 25%, 22%, 26%, 26% and 27% compared with the Strong approach.

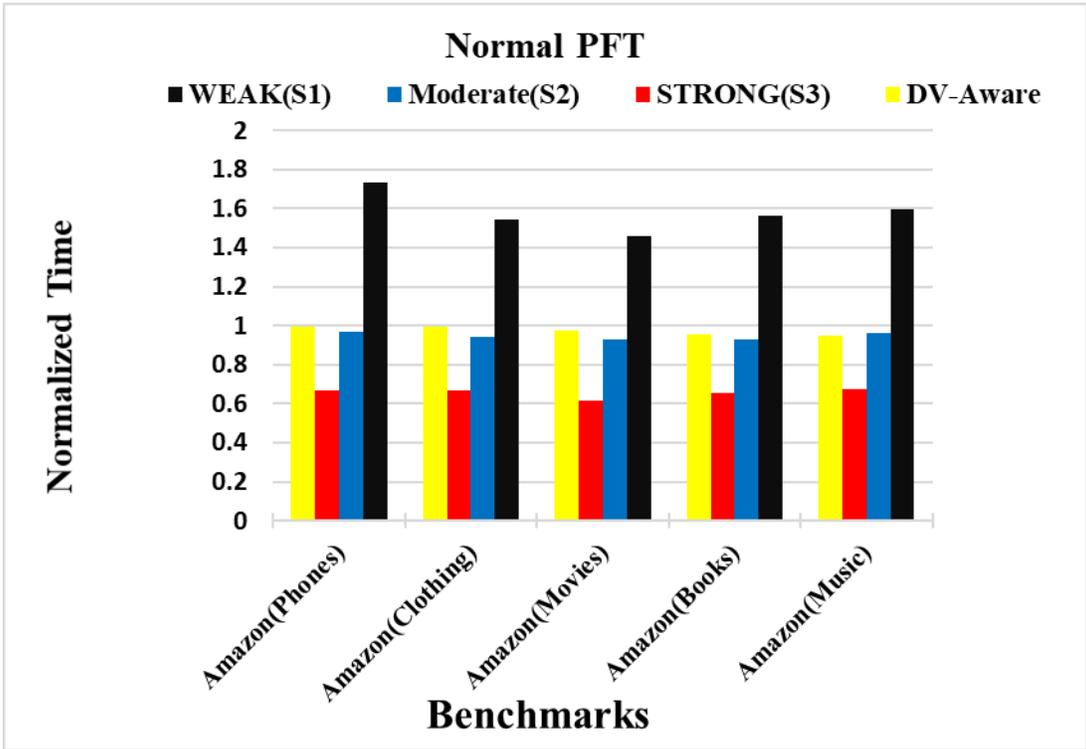

Fig. 12. Normalized Time in Normal SLO condition (Amazon Benchmarks)

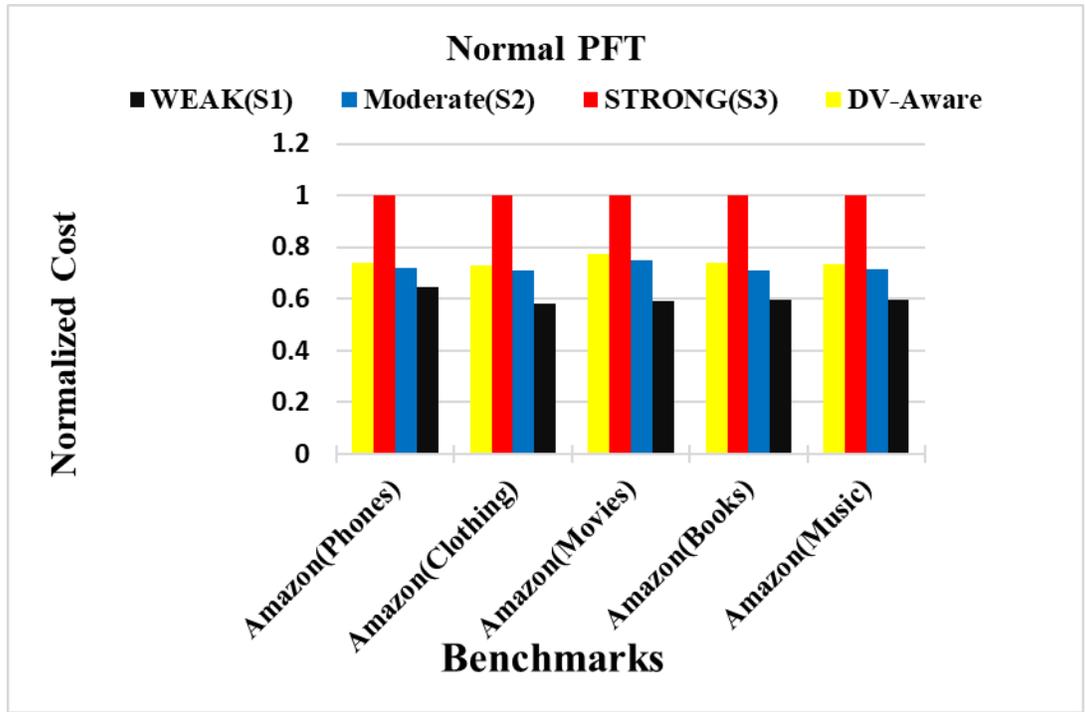

Fig. 13. Normalized Cost in Strict SLO condition (Amazon Benchmarks)

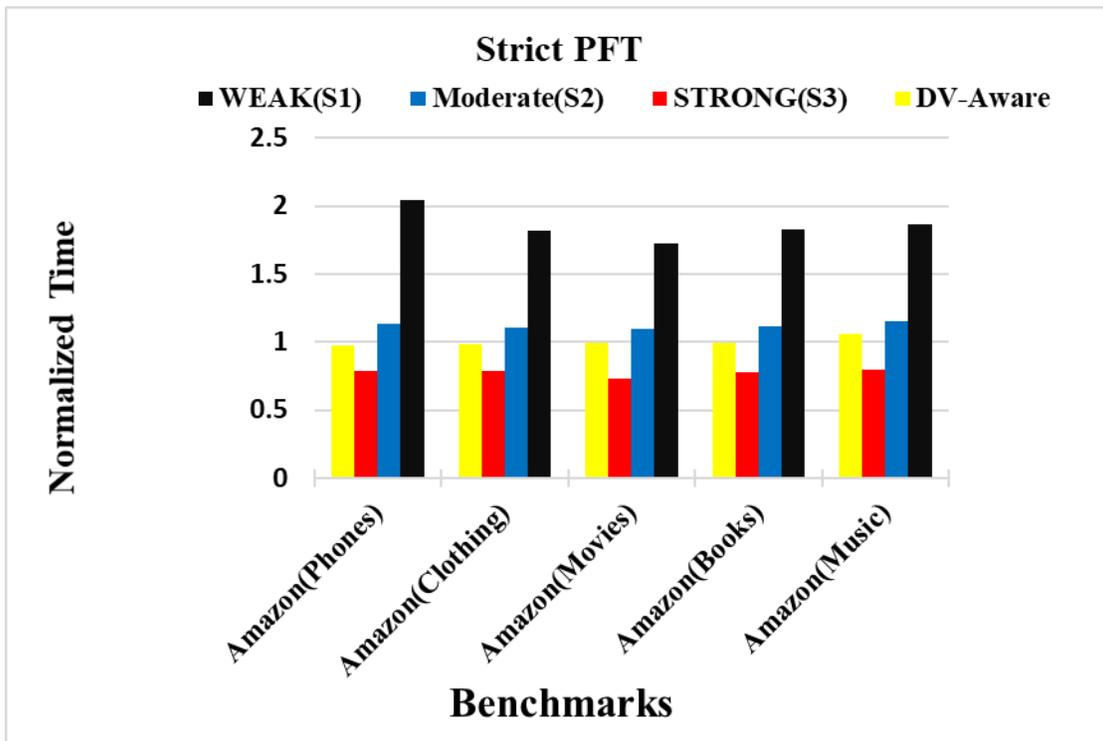

Fig. 14. Normalized Time in Normal SLO condition (Amazon Benchmarks)

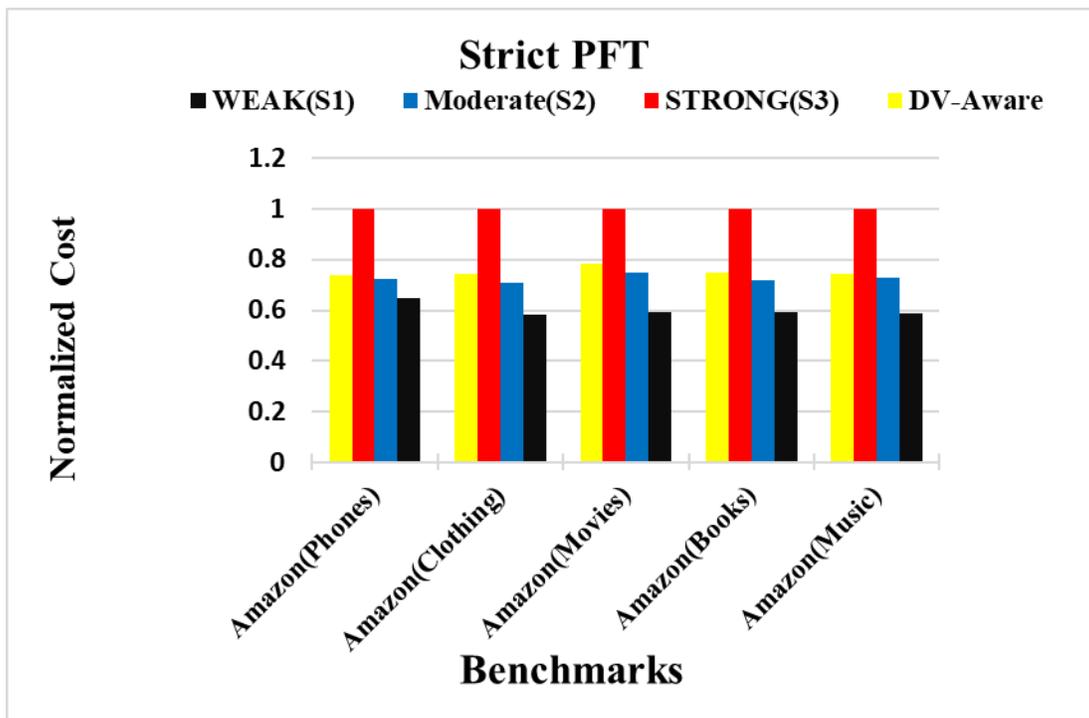

Fig. 15. Normalized Cost in Strict SLO condition (Amazon Benchmarks)

3.2 DISCUSSION

Table 5 shows the server types used for processing in each application in our approach. As Table 5 shows, more expensive servers are used in the Strict condition.

Table 5. Server types used for processing

| Application | Normal SLO condition | | | | | Strict SLO condition | | | | |
|---|---|---|---|---|---|---|---|---|---|---|
| | S1 | S2 | S3 | S4 | S5 | S1 | S2 | S3 | S4 | S5 |
| WordCount | * | * | * | | | * | | * | * | |
| Grep | * | * | * | | | * | * | | * | |
| InvertedIndex | | * | * | * | | | * | | * | * |
| Health | * | * | | * | | * | | * | | * |
| URL | * | * | * | | | * | * | | * | |
| Investment | * | * | | * | | * | | * | * | |
| TPC-H(MAIL) | * | * | * | | | * | * | | * | |
| TPC-H(SHIP) | * | * | | * | | * | | * | | * |
| TPC-H(AIR) | * | * | * | | | * | | * | * | |
| TPC-H(RAIL) | * | * | * | | | | * | * | * | |
| TPC-H(TRUCK) | * | * | * | | | * | * | | * | |
| Amazon (Music) | * | * | | * | | * | | * | * | |
| Amazon (Books) | * | * | * | | | * | * | | * | |
| Amazon (Movies) | * | * | | * | | * | * | | | * |
| Amazon (Clothing) | * | * | * | | | * | * | | * | |
| Amazon (Phones) | * | * | * | | | | * | * | * | |

Based on the observations, in case of time limitations, our approach can meet the PFT and generate the lowest processing cost in compare to data-variety-oblivious approaches.

**Functionality.** Our approach can help the users in the mentioned restrictions. In both normal and strict conditions, our approach presents a solution for server selection.

**Overheads**. The overhead of sampling and combining RDDs in less than 1%. Based on the functionality of the accumulative applications, the final result can be generated by adding the partial results. This issue can be achieved by using functions such as *join* or *union* in the Scala language.

In general, default provisioning approaches use sampling for detecting the suitable configuration for Big Data Processing. However, data variety causes differences in the significance of different parts of data. Our approach, considering data variety, divides the input data in some blocks, then uses a suitable sampling method to declare the significance of each data block, and finally assigns data blocks to suitable server configurations. Our approach also does not have more sampling overhead in compare to other approaches.

We also present
Table 6 as the verification table. In this table the time unit is second and the cost is reported relatively.

Table 6. Verification Table (Investment, URL, Health, Grep, Inverted Index and WordCount)

| | | Strict condition | | | | Normal condition | | | |
|---|---|---|---|---|---|---|---|---|---|
| | | DV-aware | STRONG(S3) | MODERATE(S2) | WEAK(S1) | DV-aware | STRONG(S3) | MODERATE(S2) | WEAK(S1) |
| **WC** | Time(s) | 34126 | 27200 | 38928 | 64865 | 37561 | 27200 | 38928 | 64865 |
| | Cost(relative) | 89512 | 108800 | 77840 | 64865 | 76821 | 108800 | 77856 | 64865 |
| **II** | Time(s) | 7191243 | 5323721 | 7761351 | 13312781 | 7619475 | 5323721 | 7761351 | 13312781 |
| | Cost(relative) | 18565345 | 21294884 | 15522702 | 13312781 | 13817112 | 21294884 | 15522702 | 13312781 |
| **GR** | Time(s) | 17953 | 13630 | 19385 | 31765 | 19257 | 13630 | 19385 | 31765 |
| | Cost(relative) | 39895 | 54520 | 38770 | 31765 | 37645 | 54520 | 38770 | 31765 |
| **HEALTH** | Time(s) | 19953 | 15630 | 22585 | 35765 | 21457 | 15630 | 22585 | 35765 |
| | Cost(relative) | 51742 | 62520 | 45170 | 35765 | 43445 | 62520 | 45170 | 35765 |
| **URL** | Time(s) | 15953 | 11930 | 18985 | 29765 | 16057 | 11930 | 18985 | 29765 |
| | Cost(relative) | 37187 | 47720 | 37970 | 29765 | 32695 | 47720 | 37970 | 29765 |
| **INVESTMENT** | Time(s) | 20953 | 16630 | 24385 | 38765 | 21957 | 16630 | 24385 | 38765 |
| | Cost(relative) | 54895 | 66520 | 48770 | 38765 | 47645 | 66520 | 48770 | 38765 |

### Table 7. Verification Table (TPC Benchmarks)

| | | Strict condition | | | | Normal condition | | | |
|---|---|---|---|---|---|---|---|---|---|
| | | DV-aware | STRONG(S3) | MODERATE(S2) | WEAK(S1) | DV-aware | STRONG(S3) | MODERATE(S2) | WEAK(S1) |
| TPC-H(MAIL) | Time(s) | 17908.12 | 13869.89 | 21308.81 | 32414.28 | 19958.44 | 13869.89 | 21308.81 | 32414.28 |
| | Cost(relative) | 41833.90 | 55479.55 | 42617.61 | 32414.28 | 38344.59 | 55479.55 | 42617.61 | 32414.28 |
| TPC-H(SHIP) | Time(s) | 17870.42 | 14817.66 | 21469.78 | 34051.67 | 20633.95 | 14817.66 | 21469.78 | 34051.67 |
| | Cost(relative) | 43686.54 | 59270.65 | 42939.56 | 34051.67 | 42357.76 | 59270.65 | 42939.56 | 34051.67 |
| TPC-H(AIR) | Time(s) | 17842.14 | 15488.04 | 21508.01 | 35762.64 | 20572.54 | 15488.04 | 21508.01 | 35762.64 |
| | Cost(relative) | 47980.92 | 61952.17 | 43016.02 | 35762.64 | 42734.60 | 61952.17 | 43016.02 | 35762.64 |
| TPC-H(RAIL) | Time(s) | 18907.20 | 14486.81 | 21391.30 | 34720.03 | 20961.48 | 14486.81 | 21391.30 | 34720.03 |
| | Cost(relative) | 48407.80 | 57947.24 | 42782.61 | 34720.03 | 41763.36 | 57947.24 | 42782.61 | 34720.03 |
| TPC-H(TRUCK) | Time(s) | 17474.55 | 15343.56 | 20839.97 | 35555.45 | 20545.32 | 15343.56 | 20839.97 | 35555.45 |
| | Cost(relative) | 45155.00 | 61374.25 | 41679.94 | 35555.45 | 39626.63 | 61374.25 | 41679.94 | 35555.45 |

### Table 8. Verification Table (Amazon Benchmarks)

| | | Strict condition | | | | Normal condition | | | |
|---|---|---|---|---|---|---|---|---|---|
| | | DV-aware | STRONG(S3) | MODERATE(S2) | WEAK(S1) | DV-aware | STRONG(S3) | MODERATE(S2) | WEAK(S1) |
| Amazon (Music) | Time(s) | 17949.59 | 13887.27 | 21004.36 | 33184.26 | 20214.12 | 13887.27 | 21004.36 | 33184.26 |
| | Cost(relative) | 41772.26 | 55549.07 | 42008.73 | 33184.26 | 39633.97 | 55549.07 | 42008.73 | 33184.26 |
| Amazon (Books) | Time | 17854.62 | 13054.03 | 20584.28 | 31193.20 | 20697.0866 | 13054.03028 | 20584.27745 | 31193.20459 |
| | Cost | 41145.68 | 52216.12 | 41168.55 | 31193.20 | 39039.46397 | 52216.1211 | 41168.5549 | 31193.20459 |
| Amazon (Movies) | Time | 17771.04 | 14096.36 | 19968.10 | 32730.88 | 21089.50 | 14096.36 | 19968.10 | 32730.88 |
| | Cost | 41899.48 | 56385.42 | 39936.20 | 32730.88 | 38652.00 | 56385.42 | 39936.20 | 32730.88 |
| Amazon (Clothing) | Time | 17474.73 | 14182.13 | 20467.30 | 36733.94 | 21089.50 | 14182.13 | 20467.30 | 36733.94 |
| | Cost | 41899.48 | 56728.54 | 40934.61 | 36733.94 | 40114.51 | 56728.54 | 40934.61 | 36733.94 |
| Amazon (Phones) | Time | 17645.68 | 14167.84 | 20993.34 | 37103.97 | 21004.49 | 14167.84 | 20993.34 | 37103.97 |
| | Cost | 41284.52 | 56671.35 | 41986.68 | 37103.97 | 41060.80 | 56671.35 | 41986.68 | 37103.97 |

5. CONCLUSIONS

In this work, we analyze the effect of Data Variety on the performance of Big Data processing. We offer a data-variety-aware approach for decreasing the Processing Cost. In this approach, we define some Data Types and process them with cost/ time efficient servers. The Data Types can be processed in parallel if enough resources are available. We evaluated our work with data-variety-oblivious approaches and showed that our data-variety-aware approach can significantly surpass other approaches. Due to the low overhead of this approach, it is applicable for the user to reduce their Processing Cost. Cloud providers can also use this approach to increase their market share. The maximum overhead of this approach is lower than 1%.

In this paper, we considered accumulative applications. In the accumulative applications, the final results are generated as the summation of partial results. In Graph Applications different parts of input data have dependency to each other, therefore considering the Graph applications can be a subject of study for the future. Analysing the impact of Data Variety on other parameters such as bandwidth or memory usage can also be another direction for feature works.